# Direct Observation of Electron Trapping States in $SiO_2$ by Constant-Final-State Photoelectron Yield Spectroscopy: Insights into the Electron-Transfer Model of Contact Electrification

Renjiro Shibuya,[1] Manato Tateno,[1] Runa Hoshikawa,[1] Tomohiro Ishibashi,[1] Hirohiko Fukagawa[1,2,3] and Hisao Ishii[1,2,3,a)]

## AFFILIATIONS

[1]Graduate School of Science and Engineering, Chiba University, 1-33 Yayoi-cho, Inage-ku, Chiba 263-8522, Japan

[2]Center for Frontier Science, Chiba University, 1-33 Yayoi-cho, Inage-ku, Chiba 263-8522, Japan

[3]Molecular Chirality Research Center, Chiba University, 1-33 Yayoi-cho, Inage-ku, Chiba 263-8522, Japan

[a)] corresponding author: ishii130@faculty.chiba-u.jp

## ABSTRACT

**Understanding electrostatic charging of silicon dioxide ($SiO_2$) is important for electrostatic control in a wide range of applications. Contact electrification (CE) is a major source of such charging, and electron transfer is considered one of its leading mechanisms. Several electron-transfer models have been proposed to explain CE; however, the density of states (DOS) involved in electron transfer has not been experimentally characterized. Determining this DOS is therefore essential for understanding the charging mechanism. However, measuring the DOS of insulating materials such as $SiO_2$ remains challenging: photoemission spectroscopy, the standard technique for probing the DOS of materials, is unsuitable for highly insulating materials because sample charging during measurement can cause spectral distortion or even complete signal loss. Here, we directly and quantitatively determined the in-gap DOS of amorphous $SiO_2$ using constant-final-state photoelectron yield spectroscopy (CFS-YS). Weak occupied states were observed within the band gap, and their marked increase after electron injection revealed that electron-trapping states are distributed over a broad energy range. Vacuum annealing markedly altered the in-gap DOS and electron-trapping behavior. The observed changes suggest a role for surface adsorbates, including water, in the charging behavior of $SiO_2$. Charging simulations based on the experimentally measured in-gap DOS approximately reproduced the previously reported charging behavior and showed that the accumulated charge was concentrated near the surface. These findings indicate that in-gap states distributed over a broad energy range mediate electron transfer during CE and highlight the importance of considering experimentally determined DOS when evaluating proposed electron-transfer models.**

Silicon dioxide ($SiO_2$) is widely used as an insulating material in electronic devices, optical fibers, and powder engineering. However, electron traps in $SiO_2$ can cause dielectric breakdown and impair device performance.[1] $SiO_2$ charging can also cause electrostatic discharge and powder aggregation. Electron traps are thought to play an important role in these charging phenomena.[2,3] Understanding the mechanisms of electron trapping and charging in $SiO_2$ is therefore essential for ensuring device reliability and controlling electrostatic charging.

Contact electrification (CE), the transfer of charge between two materials upon contact, has long been recognized as a major cause of electrostatic charging.[4] Proposed explanation models for CE include material transfer,[5] ion transfer,[6] and electron transfer.[4] However, the underlying mechanism remains unclear because CE depends on multiple factors, including humidity, surface conditions, and how the materials contact each other.[7]

The electron-transfer model is widely regarded as a leading explanation for CE. However, the wide band gaps of insulators such as $SiO_2$ mean that electron transfer between materials through their valence or conduction bands requires energies far exceeding thermal energy, making such transfer highly improbable. Several models have been proposed to explain electron transfer despite this constraint. Xu et al. proposed the electron-cloud-potential-well model, in which two approaching atoms have overlapping electron clouds that lower the potential barrier and enable electron transfer.[8] In contrast, Lowell and Rose-Innes suggested that surface states within the band gaps of insulators may participate in electron transfer during CE.[4] However, neither approach specifies the energy levels involved, limiting detailed analysis. Direct observation of the in-gap density of states (DOS) is therefore essential to identify the electronic states involved in electron transfer and determine their energy distribution.

Ultraviolet photoelectron spectroscopy (UPS) is widely used to measure occupied DOS. However, measurements of insulating materials, particularly thick films, are challenging because sample charging can distort spectra or cause complete signal loss. Nevertheless, Ohta et al. successfully measured 2.1–32.8-nm-thick $SiO_2$ films by UPS at excitation energies of 6.5–10.5 eV and evaluated the energy band diagram of the $SiO_2$/Si system. They attributed the signals from the 32.8-nm-thick film mainly to filled electronic defect states at the $SiO_2$ surface.[9] However, they did not discuss the relevance of these states to electron trapping or charging. More recently, Zhu et al. observed surface states in bulk $SiO_2$ by correcting charging-induced shifts in UPS spectra with reference to X-ray photoelectron spectroscopy (XPS) spectra. However, this method is limited by the energy resolution of XPS, and the possible spatial inconsistency in charging and the associated spectral broadening effect remain to be clarified.[10] Photoelectron yield spectroscopy (PYS) is another technique for investigating occupied DOS. Ohta et al. determined the DOS distribution of a 40-nm-thick $SiO_2$ film by differentiating its PYS spectrum.[11] Although PYS can effectively probe the electronic states of insulators,[12] estimating DOS by differentiating PYS spectra is qualitative, and discrepancies with UPS results have been reported.[13] A more quantitative method for measuring DOS is therefore required.

We previously developed high-sensitivity ultraviolet photoelectron spectroscopy (HS-UPS), a modification of conventional UPS that uses low-energy photons free of stray photons to reduce the background signal.[14] HS-UPS can detect photoelectron emission from in-gap states. Low-energy photons suppress sample charging and photoinduced damage by substantially reducing primary and secondary electron emission compared with conventional UPS.[15] Constant -final-state photoelectron yield spectroscopy (CFS-YS),[16,17] which measures photoelectron yield at a fixed kinetic energy while varying photon energy, detects weak spectral features with a higher signal-to-noise ratio than HS-UPS.[13,15] It can also probe occupied states associated with electron traps. Conventional trap spectroscopy generally measures trap depths relative to the conduction-band minimum (CBM), whereas CFS-YS determines the absolute energies of these states relative to the vacuum level.

In this study, we used CFS-YS to directly observe in-gap states in $SiO_2$. We then induced negative charging by electron-beam irradiation to examine occupied states associated with electron traps. We examined how annealing affected the DOS and charging characteristics, focusing on the desorption of water-related surface species. Finally, we used the DOS measured by CFS-YS in charging simulations to evaluate the amount and spatial distribution of accumulated charge. These investigations clarify the roles of in-gap states in the CE of $SiO_2$ and enable assessment of the electron-transfer model.

The sample was a 100-nm-thick thermally grown $SiO_2$ film on a p-type Si(100) substrate (≤0.02 Ω cm; Selen Advanced Materials Corp.). It was ultrasonically cleaned with acetone and isopropanol, then dried with nitrogen gas. The sample was annealed under vacuum (~$10^{-5}$ Pa) at 300 °C for 1 h using an infrared heater (GVH298, Thermo Riko Co. Ltd.). Electron-beam irradiation was achieved by operating an ion gauge for 10 min with a positive bias applied to the sample. For HS-UPS, monochromatic photons at 7.7 eV irradiated the sample, and an electrostatic hemispherical electron analyzer (PSP RESOLVE120) measured the kinetic energy distribution of emitted photoelectrons. The same setup was used for CFS-YS. Further details of the measurement conditions and experimental setup are provided in the supplementary material.

The charging simulations calculated charge transfer at the metal/$SiO_2$ interface under thermal equilibrium using the occupied DOS measured by CFS-YS. The hole density was estimated by applying the Fermi–Dirac distribution function to this DOS. Figure 1 illustrates the charge-transfer framework, which adopts the charge neutrality level (CNL) of $SiO_2$ as the charge-neutral reference, following our previous study.[15] Since CFS-YS probes only occupied states, we introduced a CNL-based correction term to perform the calculations without the unoccupied-state DOS. Details are provided in the supplementary material.

Figure 2 compares the HS-UPS spectrum of a 100-nm-thick $SiO_2$ film on p-type Si(100), measured using 7.7-eV photons (red), with the He I UPS (black, 21.2 eV) and metastable atom electron spectroscopy (MAES, blue, 19.8 eV) spectra of $SiO_x$ (~0.6 nm) on n-type Si(111) reported by Ishii et al.[18] All spectra are plotted against binding energy relative to the vacuum level. For He I UPS, an ultrathin $SiO_x$ film was used to avoid severe sample charging. However, because the oxide thickness is comparable to the UPS probing depth,[19] signals from bulk Si are also detected, as indicated by the hatched region in Figure 2, complicating selective evaluation of the oxide layer's electronic states. In contrast, metastable He atoms cannot penetrate the bulk, allowing MAES to selectively probe the electronic states of the outermost oxide layer without interference from the underlying substrate. However, the MAES spectrum in Figure 2 exhibits no clear in-gap states, and, to our knowledge, such states in $SiO_2$ have not yet been experimentally observed by MAES. By comparison, HS-UPS suppressed sample charging even for the 100-nm-thick film and clearly detected weak in-gap states in $SiO_2$. Although similar states were previously reported, their relevance to electron trapping and charging remained unclear.[9] We therefore investigated their roles in these processes using CFS-YS.

Figure 3 shows the CFS-YS spectra before and after electron-beam irradiation, which negatively charged the sample and populated electron traps. The CFS-YS intensity was converted to DOS as described in the supplementary material. For the pristine sample (black), weak occupied states appeared at an onset energy of approximately 4 eV and extended toward the valence-band maximum (VBM, ~10 eV)[18,20], indicating a broad distribution of occupied states within the band gap. This distribution was resolved over a dynamic range of approximately seven orders of magnitude. For comparison, Figure 3 also shows the DOS of pristine $SiO_2$ particles (open circles; see the supplementary material for sample preparation). The particles have a high surface-to-bulk ratio and exhibit a higher in-gap DOS than the film, indicating a substantial surface contribution to the observed in-gap states. This particle–film comparison is discussed further alongside the simulation results. We then compared the CFS-YS spectrum with the differentiated PYS (DPYS) spectra reported by Ohta et al. to examine their spectral characteristics and detection limits (Figures S2(a) and S2(b) in the supplementary material).[11] Since the DPYS spectra used a VBM-referenced energy scale, the CFS-YS energy axis was aligned either at the spectral onset [Figure S2(a)] or by assuming a $SiO_2$ VBM of 10.3 eV[18] [Figure S2(b)]. The CFS-YS and DPYS spectra differed in overall shape, possibly owing to differences in sample surface conditions and the qualitative nature of DOS estimation from DPYS. The CFS-YS detection limit was approximately two and four orders of magnitude lower than that of DPYS in Figures S2(a) and S2(b), respectively, demonstrating the higher sensitivity of CFS-YS.

Prior to negative charging, the sample was annealed in vacuum at 300 °C for 1 h to clean the surface. After annealing (red), the onset shifted to a lower binding energy, from approximately 4 to 2 eV, and the DOS increased over a broad energy range. These changes may reflect the removal of physisorbed water and partial surface dehydroxylation.[21] The effects of annealing on the DOS and charging behavior are discussed in detail later. Additionally, as indicated by the arrow in Figure 3, the onset lies above the substrate Fermi level, suggesting that the sample is not in thermal equilibrium.

After electron-beam irradiation at a sample bias of $V_s$ = 100 V (blue), the intensity increased over approximately 2–5 eV. At $V_s$ = 200 V (purple), a further increase occurred at binding energies of approximately 3–8 eV, where no pronounced increase was observed at 100 V. These increases indicate that injected electrons were trapped in in-gap states. The increased signal partially decayed over time, consistent with electron detrapping (Figure S3 of the supplementary material), indicating that the increase was unlikely to result from irreversible defect generation by electron irradiation.

To explore the origins of these electron-trapping states, we compared their energy ranges with previously reported trap levels. Reported electron traps include a broad distribution centered at 2.1 eV below the CBM (width ~0.5 eV, attributed to local disorder),[22] along with levels at 2.8 and 3.2 eV below the CBM associated with excess silicon[23] and Si sites with wide O–Si–O bond angles,[24] respectively. The intensity increases after irradiation at 100 V broadly overlapped these reported trap-energy ranges. Notably, the additional intensity at $V_s$ = 200 V overlapped the theoretically predicted energies of $Si^-$ ($I_v$ = 6.02 eV) and $SiO^-$ ($I_v$ = 7.32 eV) in β-cristobalite. Unlike the corresponding paramagnetic radicals detectable by electron paramagnetic resonance (EPR), these negatively charged states have primarily been studied theoretically, with limited experimental evidence.[3] CFS-YS therefore offer experimental access to these predicted negatively charged defect states. Together, the spectral changes indicate that injected electrons can be trapped over a broader energy range within the $SiO_2$ band gap than previously reported, although the origins of these states remain unclear. The presence of in-gap states capable of retaining electrons supports their participation in electron transfer during CE.

We next examined the spectral changes induced by annealing. Previous studies have shown that vacuum heating removes physisorbed water and induces partial dehydroxylation of amorphous silica surfaces through silanol condensation.[21] To assess how these surface changes affect electron trapping and charging, we performed electron-beam irradiation at $V_s$ = 100 V with and without prior annealing. As shown in Figures 4(a) and 4(b), the trapped-electron density, calculated by integrating the irradiation-induced DOS increase over 2–5 eV, was nearly two orders of magnitude higher in the annealed sample. This marked increase indicates that annealing-induced surface changes enhance electron trapping and strongly influence the charging behavior of $SiO_2$.

Finally, we performed charging simulations to assess the electron-transfer model for CE. We used the DOS measured after electron injection at $V_s$ = 200 V to account for initially empty states that become occupied during charging and set the $SiO_2$ CNL to 5.5 eV based on a previous study.[25] We then calculated the surface charge density and its spatial distribution. Figure 5(a) shows the surface charge density against the work function of the metal in contact with $SiO_2$. The open symbols represent experimental data reported by Labadz et al., [26] and the blue line represents the simulation. The two differed substantially in absolute charge density. To compare their work-function-dependent trends, we rescaled the experimental values by a constant factor of seven. The scaled values (red symbols) agreed reasonably with the simulation, indicating that the in-gap DOS can capture the observed charging trend. The difference in absolute charge density may reflect differences in experimental conditions, the actual contact area, and uncertainty in the absolute DOS used in the simulation. Figure 5(b) shows charge-density profiles in $SiO_2$ in contact with metals of different work functions. The estimated charge penetration depth was approximately 10 nm, consistent with the ~30 nm upper limit reported for Mg/$SiO_2$ contacts by Labadz et al.,[26] suggesting that charge accumulates near the $SiO_2$ surface. Together with the substantial surface contribution to in-gap states inferred from the particle–film comparison in Figure 3, these results indicate that electron transfer during CE proceeds through in-gap states near the $SiO_2$ surface. However, the simulations assumed thermal equilibrium, whereas actual CE may involve nonequilibrium processes. Incorporating these processes into future simulations may clarify the charging mechanism further.

In conclusion, we directly observed occupied in-gap states in $SiO_2$ using CFS-YS. Electron-injection experiments revealed a broad distribution of occupied states associated with electron traps within the band gap. Vacuum annealing substantially increased the trapped-electron density, suggesting that water-related surface species, including physisorbed water and hydroxyl groups, may influence electron trapping and charging in $SiO_2$. Charging simulations based on the experimentally measured in-gap DOS approximately reproduced the observed charging trend and showed that accumulated charge was predominantly localized near the surface. These findings indicate that in-gap states participate in electron transfer and trapping during CE, contributing to charging and providing experimental support for the electron-transfer model. They also highlight the importance of using experimentally determined DOS to evaluate such models. Beyond CE, these insights may help explain other electron-trap-related phenomena, including device degradation and dielectric breakdown.

## SUPPLEMENTARY MATERIAL

See the supplementary material for additional details on the experimental setup, simulation model, DOS conversion procedure, comparison with previous studies, time-dependent electron detrapping, and preparation of $SiO_2$ particle samples.

**ACKNOWLEDGMENT**

We thank Prof. Scott Waitukaitis and Dr. Andrea Stöllner of ISTA for helpful discussions on the $SiO_2$ particle samples.

**AUTHOR DECLARATIONS**

**Conflict of Interest**

The authors have no conflicts to disclose.

**Author Contributions**

**Renjiro Shibuya**: Conceptualization (equal); Data curation (lead); Formal analysis (lead); Investigation (lead); Writing – original draft (lead); Writing – review & editing (equal). **Manato Tateno**: Conceptualization (equal); Investigation (equal). **Runa Hoshikawa**: Software (lead); **Tomohiro Ishibashi**: Methodology (equal); Resources (equal). **Hirohiko Fukagawa**: Writing – review & editing (equal). **Hisao Ishii**: Conceptualization (lead); Funding acquisition (lead); Methodology (lead); Project administration (lead); Resources (lead); Supervision (lead); Writing – review & editing (lead).

**DATA AVAILABILITY**

The data that support the findings of this study are available from the corresponding author upon reasonable request.

## Figures

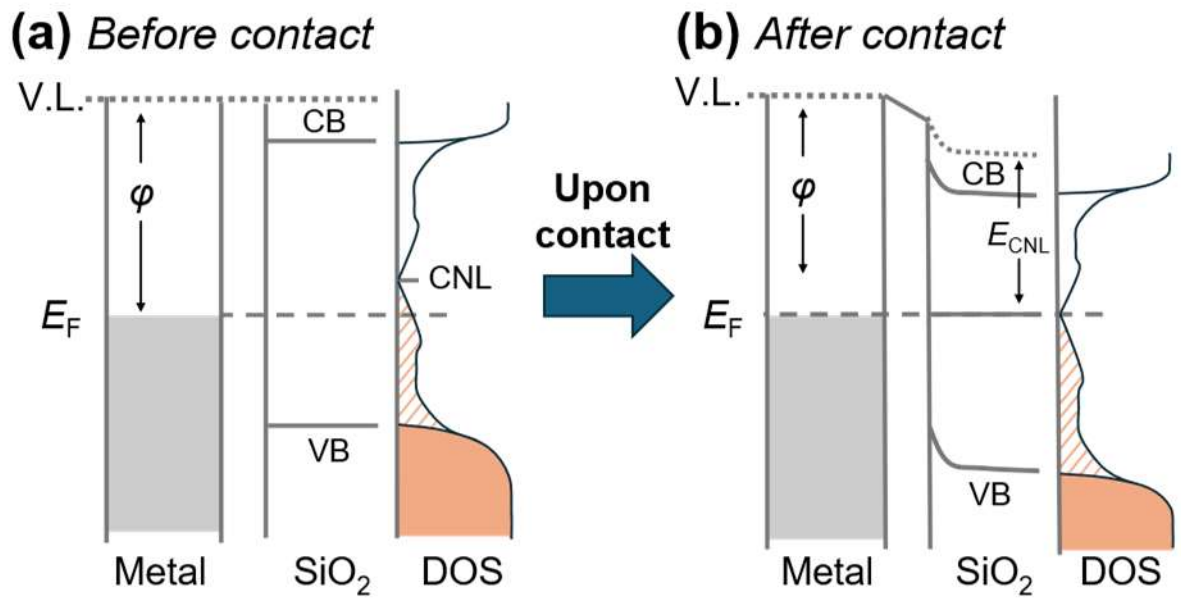

Figure 1. Schematic of the charge-transfer model used in this study, based on the framework proposed in Ref. 15: (a) before and (b) after contact between a metal and $SiO_2$. Upon contact, electrons transfer until the metal Fermi level aligns with the $SiO_2$ CNL. Here, $\varphi$, $E_F$, and $E_{CNL}$ denote the metal work function, Fermi level, and CNL energy, respectively; CB and VB denote the conduction and valence bands.

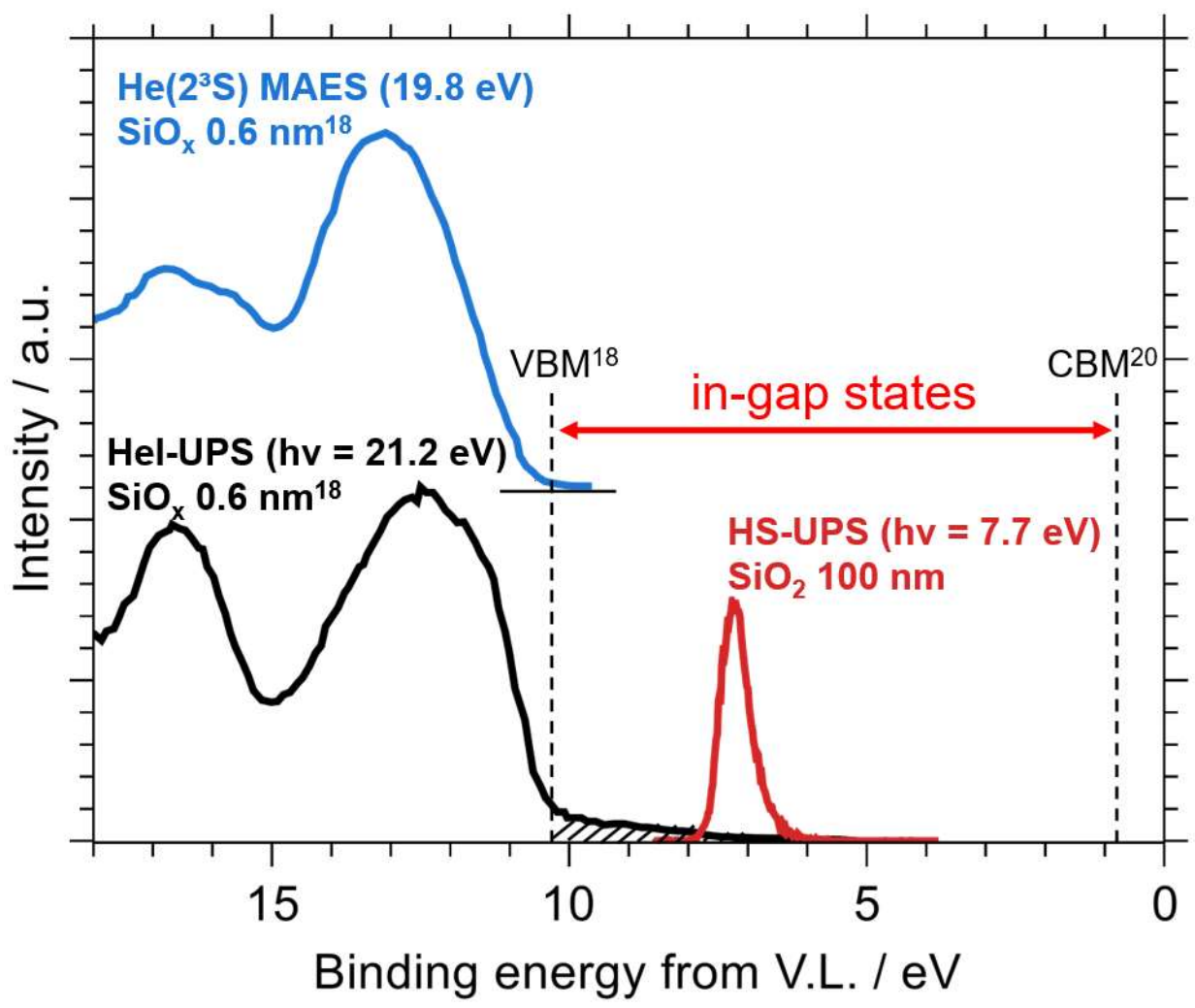

Figure 2. Comparison of the He I UPS (black, 21.2 eV) and He*($2^3S$) MAES (blue, 19.8 eV) spectra of $SiO_x$ (~0.6 nm)/n-type Si(111) reported by Ishii et al.[18] with the HS-UPS spectrum (red, 7.7 eV) of a 100-nm-thick $SiO_2$ film on p-type Si(100) measured in this study. The valence-band maximum (VBM) and conduction-band minimum (CBM) were taken from Refs. 18 and 20, respectively.

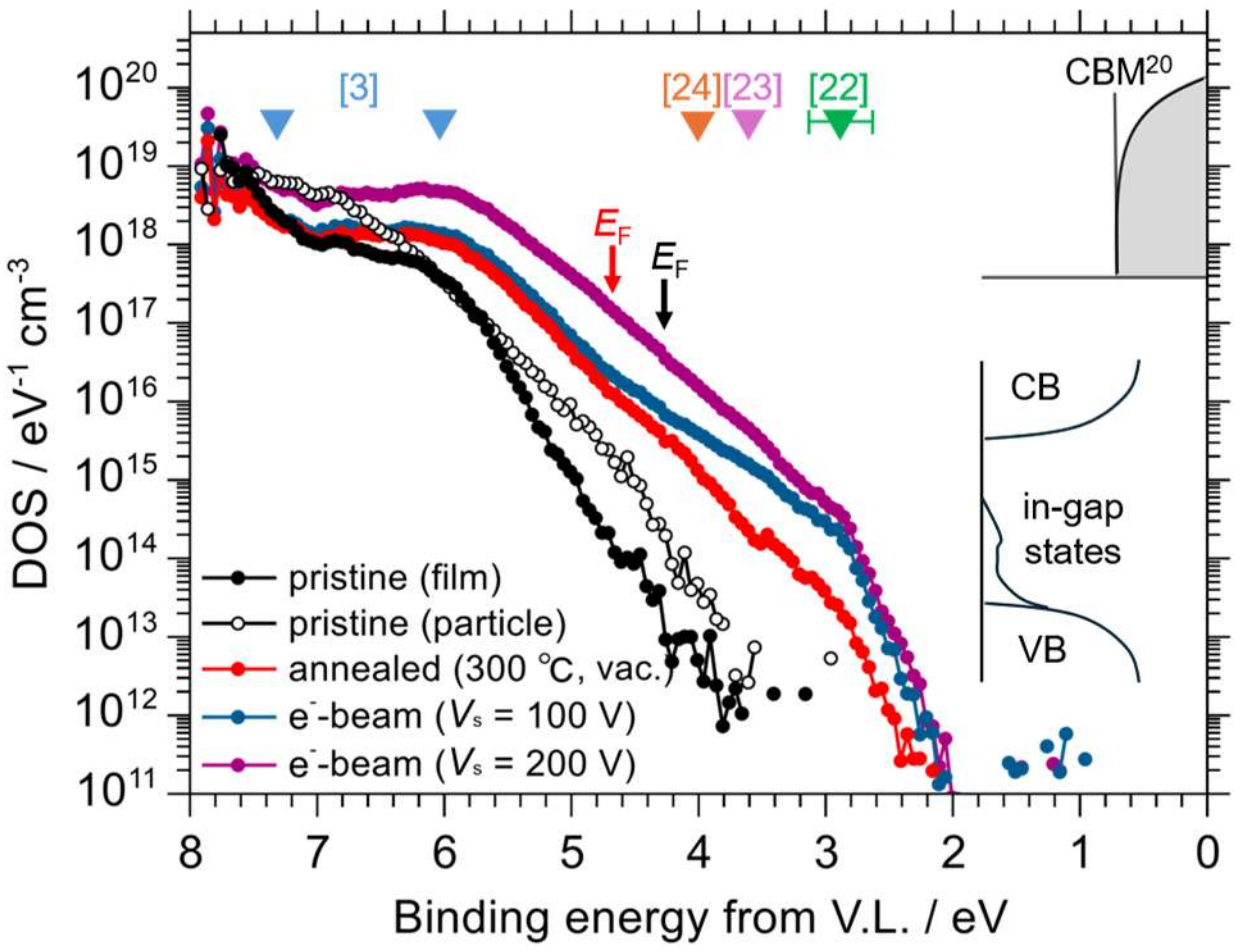

Figure 3. CFS-YS spectra of a 100-nm-thick $SiO_2$ film on p-type Si(100). The black, red, blue, and purple curves correspond to the pristine sample, the sample vacuum-annealed at 300 °C for 1 h, and the samples irradiated with an electron beam at sample biases ($V_s$) of 100 and 200 V, respectively. Open circles represent the DOS of pristine $SiO_2$ particles, normalized to the pristine film spectrum in the high-binding-energy region. Black and red arrows indicate the substrate Fermi levels before and after annealing, respectively. Inverted triangles indicate reported electron-trap energy levels in $SiO_2$.[3,22–24]

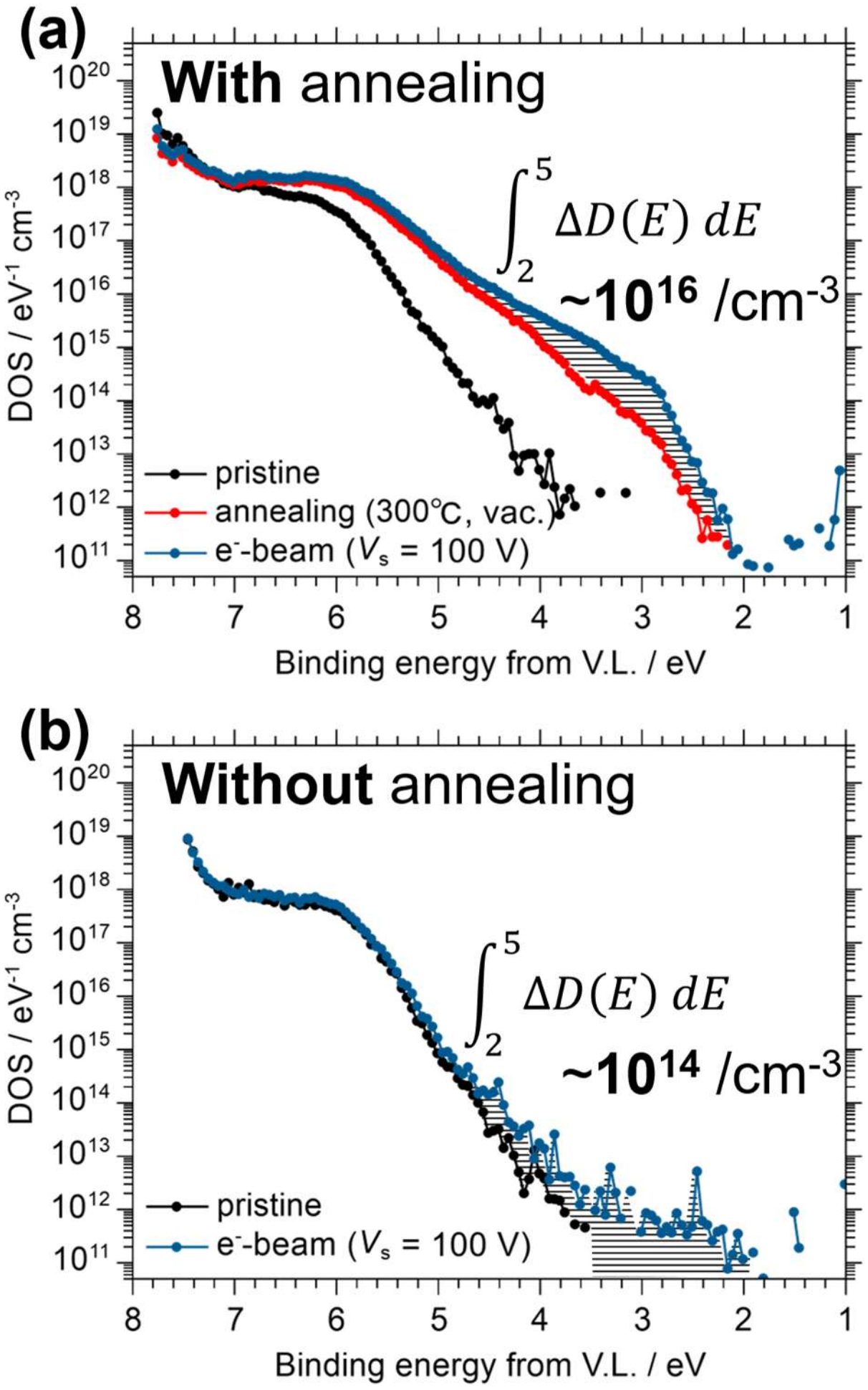

Figure 4. Comparison of trapped-electron distributions after electron-beam irradiation ($V_s$ = 100 V): (a) with and (b) without prior annealing. The hatched region indicates the DOS integrated over 2–5 eV.

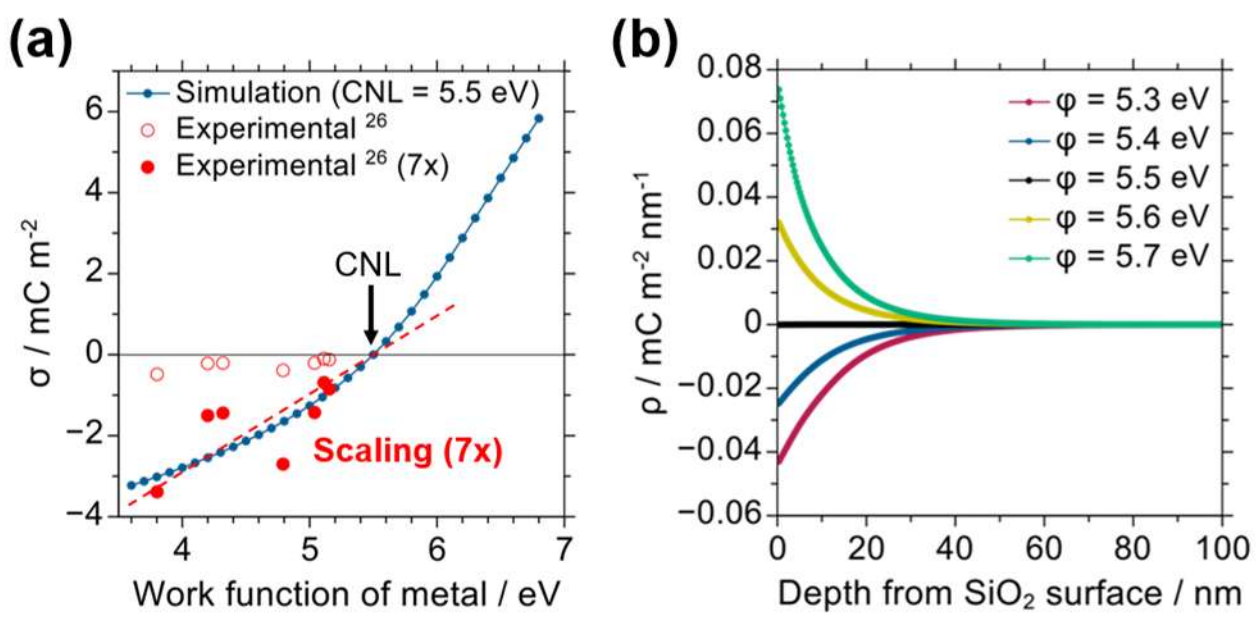

Figure 5. (a) Simulated surface charge density as a function of the work function of the metal in contact with $SiO_2$. Open and filled circles represent the experimental values reported by Labadz et al.[26] before and after scaling by a factor of seven, respectively. The Au-referenced work-function scale in Ref. 26 was converted to an absolute scale by assuming an Au work function of 5.1 eV.[27] (b) Simulated charge-density profiles in $SiO_2$ in contact with metals of different work functions, plotted against depth from the $SiO_2$ surface.

Supplementary Material

**Direct Observation of Electron Trapping States in $SiO_2$ by Constant-Final-State Photoelectron Yield Spectroscopy: Insights into the Electron-Transfer Model of Contact Electrification**

Renjiro Shibuya,[1] Manato Tateno,[1] Runa Hoshikawa,[1] Tomohiro Ishibashi,[1] Hirohiko Fukagawa[1,2,3] and Hisao Ishii[1,2,3,a)]

**AFFILIATIONS**

[1]Graduate School of Science and Engineering, Chiba University, 1-33 Yayoi-cho, Inage-ku, Chiba 263-8522, Japan
[2]Center for Frontier Science, Chiba University, 1-33 Yayoi-cho, Inage-ku, Chiba 263-8522, Japan
[3]Molecular Chirality Research Center, Chiba University, 1-33 Yayoi-cho, Inage-ku, Chiba 263-8522, Japan

[a)] corresponding author: ishii130@faculty.chiba-u.jp

**HS-UPS and CFS-YS measurements**

Before analysis, the sample was sprayed with isopropanol to reduce residual electrostatic charge. Its surface potential was then measured using a Kelvin probe (McAllister Technical Services) to confirm that no substantial residual charge remained.

For HS-UPS, the sample was irradiated with monochromatic photons, and the kinetic energy distribution of emitted photoelectrons was measured using an electrostatic hemispherical electron analyzer (PSP RESOLVE120). A sample bias of −50 V was applied to sufficiently accelerate the emitted photoelectrons. Monochromatic light ($h\nu$ = 1.5–8.3 eV) was generated using a deuterium lamp (Hamamatsu Photonics, L1835, 150 W) and a xenon lamp (Ushio, UXL-500D, 500 W) combined with a zero-dispersion double monochromator (Bunkoukeiki, BIP-M25 GTM). Low-energy photons suppressed sample charging, enabling UPS measurements of insulating materials such as $SiO_2$.

CFS-YS was performed using the same setup as HS-UPS. The photon energy was scanned while monitoring the photoelectron yield at a fixed kinetic energy, providing higher sensitivity to weak electronic states than HS-UPS. The kinetic energy of detected photoelectrons was fixed at 0.54 eV. The measurement principle is described in detail elsewhere.[1]

**Simulation model**

The simulation model was based on the energy-level alignment model for organic semiconductor/metal interfaces reported by Oehzelt *et al.*[2] In this model, the highest occupied molecular orbital (HOMO) and lowest unoccupied molecular orbital (LUMO) densities of states (DOS) of the organic semiconductor are approximated by Gaussian functions. The charge density and electrostatic potential distributions at the metal/organic semiconductor interface are calculated self-consistently under thermal equilibrium. The charge density and Poisson equation are given by Eqs. (1) and (2), respectively.

$$\rho(z) = -e\left\{\int_{-\infty}^{\infty} D_H[E + eV(z)] \times [1 - f_{FD}(E)]dE - \int_{-\infty}^{\infty} D_L[E + eV(z)] \times f_{FD}(E)dE\right\}. (1)$$

$$\frac{d^2V(z)}{dz^2} = -\frac{\rho(z)}{\varepsilon_r\varepsilon_0}. (2)$$

where $\rho(z)$ is the charge density; $V(z)$ is the electrostatic potential; $D_H$ and $D_L$ are the occupied and unoccupied DOS, respectively; $f_{\mathrm{FD}}$ is the Fermi–Dirac distribution function; $e$ is the elementary charge; and $\varepsilon_r$ and $\varepsilon_0$ are the relative permittivity of $SiO_2$ and the vacuum permittivity, respectively.

To apply this model, we replaced the Gaussian DOS with the occupied DOS measured by CFS-YS. Ideally, calculating the charge density requires both occupied and unoccupied DOS. However, CFS-YS cannot measure unoccupied DOS. Following Hoshikawa et al.,[3] we therefore introduced the charge neutrality level (CNL) of $SiO_2$ as the reference for the charge-neutral condition. When the Fermi level coincides with the CNL, the electron and hole densities in $SiO_2$ are assumed to be equal. We defined this electron density (equal to the hole density) as $\rho_{\mathrm{CNL}}$ and introduced it as a correction term in Eq. Eq. (3).

$$\rho(z) = -e\left\{\int_{-\infty}^{\infty} D_H[E + eV(z)] \times [1 - f_{FD}(E)]dE - \rho_{CNL}\right\}. (3)$$

where $\rho_{\mathrm{CNL}}$ is the hole density (equal to the electron density) when the Fermi level coincides with the CNL at thermal equilibrium. This treatment enables the charge density to be calculated from the occupied DOS alone.

**Particle sample preparation**

Monodisperse silica particles (Cospheric, SiO2MS-2.0) with a mean diameter of 0.69 μm and a density of 2.0 g $cm^{-3}$ were used. Before particle deposition, indium tin oxide (ITO) substrates were ultrasonically cleaned in acetone and isopropanol, then treated with UV–ozone to render their surfaces hydrophilic. A suspension containing 1.5 vol% silica was prepared in a 1:1 (vol/vol) water–ethanol mixture. The solvent composition and silica concentration were selected based on a previous study.[4] The particles were dispersed by magnetic stirring at 800 rpm and ultrasonication. Finally, 50 μL of the suspension was drop-cast onto the ITO substrate using a micropipette and dried in air.

**Conversion of CFS-YS spectra to DOS**

First, the MAES spectrum of $SiO_2$ was prepared.[5] The previously reported theoretical DOS of α-quartz was used as a reference to convert the spectral intensity to DOS.[6] The MAES intensity was integrated between the valence-band maximum (VBM) and VBM + 4.2 eV, corresponding to the full energy range of the reported theoretical DOS. The spectrum was then multiplied by a constant to match the integrated theoretical DOS over the same energy range.

Next, the CFS-YS spectrum was scaled to overlap with the DOS-converted MAES spectrum. However, the CFS-YS spectrum did not cover a sufficiently wide energy range for direct alignment. To bridge this gap, we used the PYS spectrum of thermally grown $SiO_2$ (40 nm/Si) reported by Ohta et al. , as differentiated PYS (DPYS) spectra have been reported to reflect occupied DOS.[7] The reported PYS spectrum was digitized, smoothed using a cubic smoothing spline, and differentiated to obtain the DPYS spectrum. Finally, the CFS-YS spectrum was aligned with the DOS-converted MAES spectrum through the resulting DPYS spectrum (Figure S1).

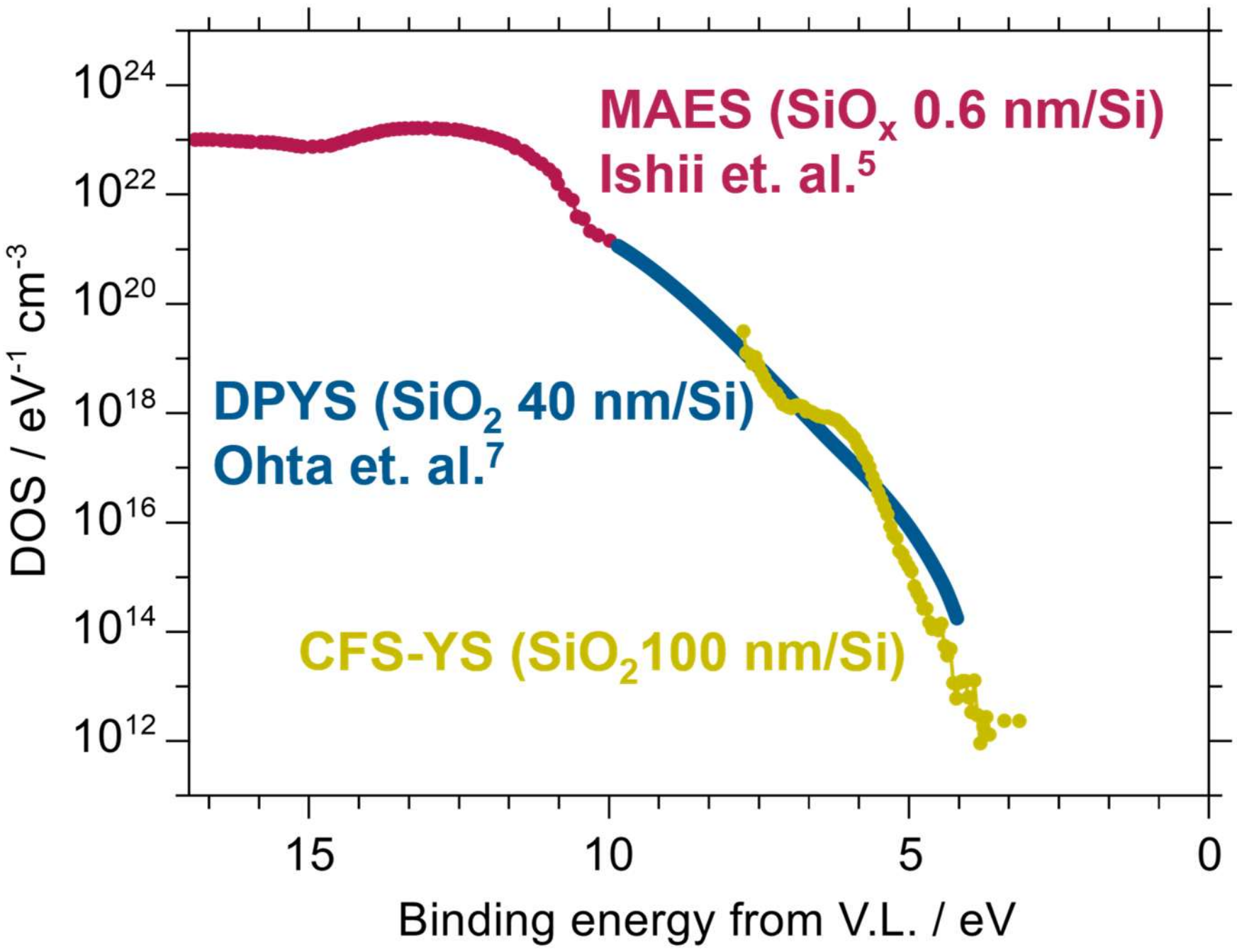

Figure S1. Conversion of the CFS-YS spectrum to DOS. The PYS spectrum of thermally grown $SiO_2$ (40 nm/Si) reported by Ohta *et al.*[7] was differentiated to obtain the DPYS spectrum and aligned with the CFS-YS spectrum. The combined spectrum was then aligned with the MAES spectrum of $SiO_x$ (0.6 nm/Si),[5] converted to DOS using the theoretical DOS of α-quartz.[6]

**Alt text:** Logarithmic DOS plot showing CFS-YS, differentiated PYS, and MAES spectra spanning complementary energy ranges. Their overlapping regions form a continuous distribution from the valence-band region to the much weaker in-gap states of $SiO_2$.

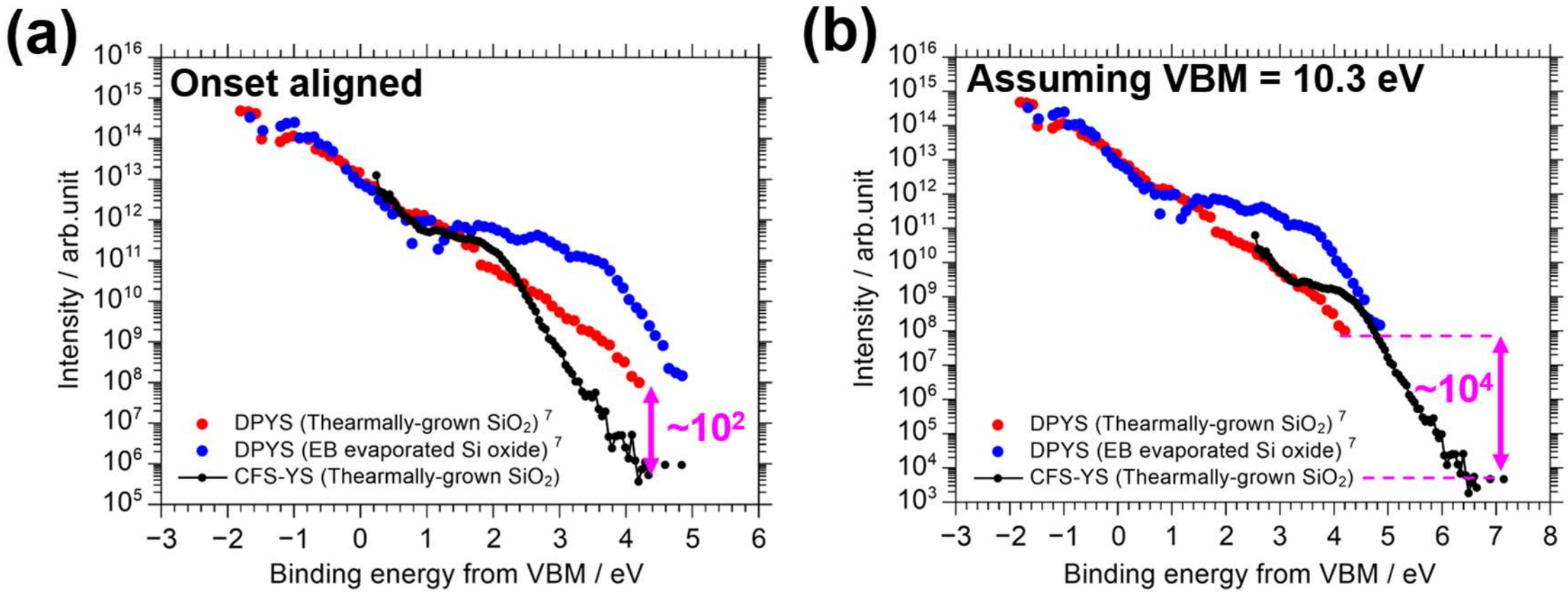

Figure S2. Comparison of the DOS measured by CFS-YS with the differentiated photoelectron yield spectroscopy (DPYS) spectra of thermally grown $SiO_2$ (red) and electron-beam (EB)-evaporated silicon oxide (blue) reported by Ohta *et al.*[7] As the DPYS spectra were referenced to the valence-band maximum (VBM), the CFS-YS spectrum is also plotted on a VBM-referenced energy scale. (a) CFS-YS spectrum shifted to align its onset with that of thermally grown $SiO_2$ reported by Ohta et al. (b) CFS-YS spectrum plotted assuming a $SiO_2$ VBM of 10.3 eV, as reported in Ref. 5. The CFS-YS detection limit is approximately two and four orders of magnitude lower than that of DPYS in panels (a) and (b), respectively.

**Alt text:** In both panels, the CFS-YS spectrum extends far below the previously reported DPYS spectra of $SiO_2$, highlighting its much higher sensitivity to weak in-gap states. The advantage is about two orders of magnitude in panel (a) and four orders in panel (b).

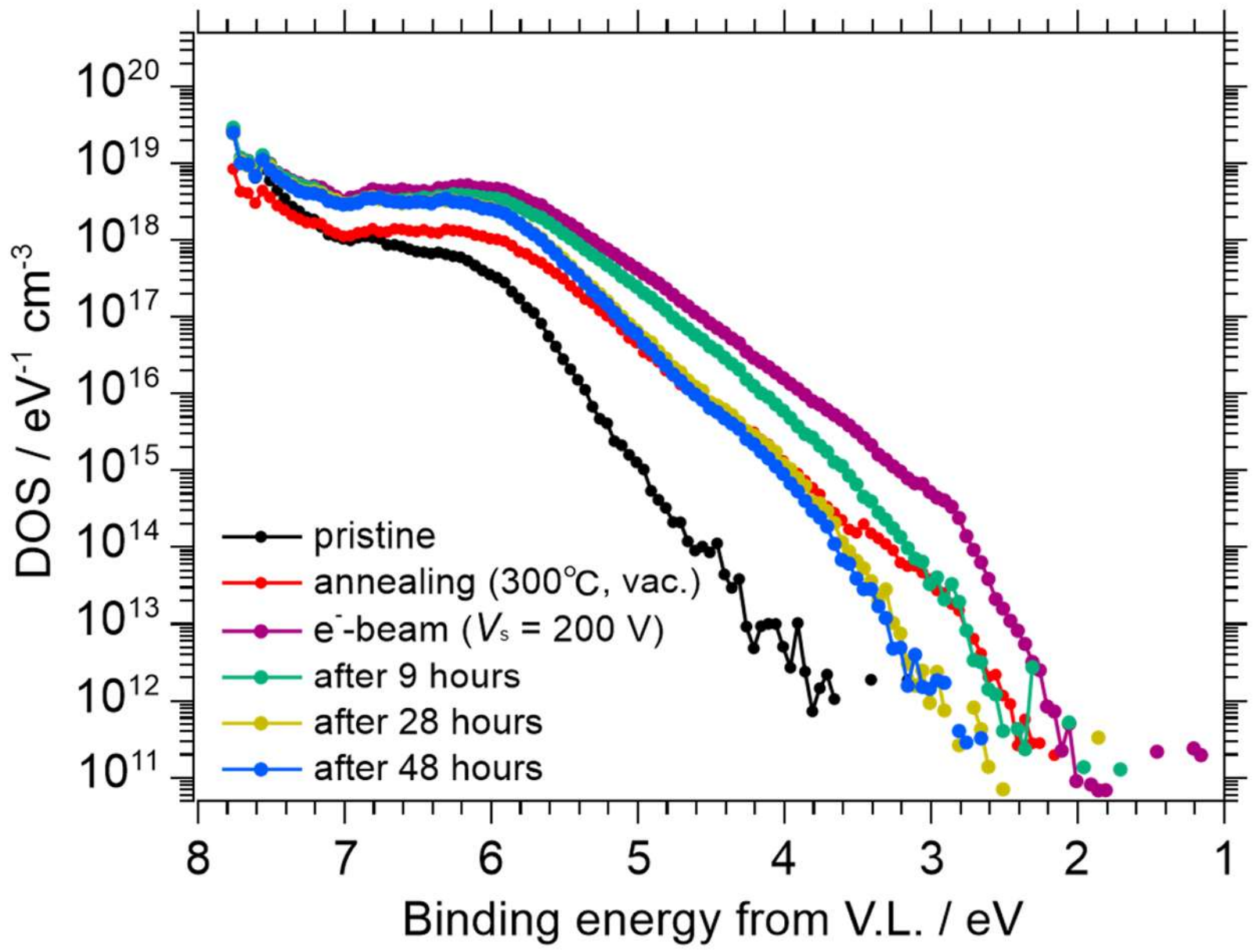

Figure S3. Time evolution of CFS-YS spectra after electron-beam irradiation at $V_s = 200$ V. The irradiation-induced occupied states gradually decreased over time, indicating electron detrapping. However, the intensity over part of the energy range fell below that of the annealed sample; the origin of this behavior remains unclear.

**Alt text:** Logarithmic DOS spectra of pristine, annealed, and electron-beam-irradiated $SiO_2$ measured at different times after irradiation. Electron irradiation increases occupied in-gap states, while the enhanced DOS gradually decreases after 9, 28, and 48 hours, indicating progressive electron detrapping.